# Analysis of Transient Processes in a Radiophysical Flow System

**E. N. Egorov*, A. A. Koronovskiĭ, and A. E. Hramov**
*State Scientific Center "College," Saratov State University, Saratov, Russia*
** e-mail: Egorov@cas.ssu.runnet.ru*

**Abstract**—Transient processes in a third-order radiophysical flow system are studied and a map of the transient process duration versus initial conditions is constructed and analyzed. The results are compared to the arrangement of submanifolds of the stable and unstable cycles in the Poincaré section of the system studied.

In most investigations devoted to various dynamical systems, the effort is concentrated on the established regimes while the transient processes preceding the attainment of a certain stable state receive much less attention. However, the knowledge about behavior of the imaging point before attaining the attractor, the duration of this transient process, and the dependence of the transient time on the control parameters provides a deeper insight into various phenomena observed in the system (such as, e.g., transient chaos—a phenomenon representing essentially the transient process of a special kind). Previously, we have studied in much detail the transient processes in discrete maps [4–6].

This study addresses transient processes in a model system with continuous time and considers dependence of the character of these processes on the regime of oscillations and on the arrangement of the manifolds of saddle cycles in the phase space.

The model system is a two-circuit radiophysical autooscillator described by the following system of equations [7]:

$$\frac{dx}{d\tau} = \frac{(\alpha-1)f(x)-z}{\gamma},$$
$$\frac{dy}{d\tau} = -\frac{\alpha f(x)}{\gamma}, \qquad (1)$$
$$\frac{dz}{d\tau} = \gamma(x+y),$$

where $\alpha$ and $\gamma$ are the control parameters and $f(x)$ is a dimensionless function determining the current–voltage characteristic of the nonlinear element of the system under consideration. This characteristic has the form

$$f(x) = -\frac{1}{2}x + \frac{3}{4}(|x+1| - |x-1|), \qquad (2)$$

representing a three-segment piecewise linear function. The system (1) has been studied in sufficient detail [9–11] and, despite simple circuitry and the form of the characteristic function (2), admits complex periodic, quasi-periodic, and chaotic oscillations in the absence of external action [8]. We will consider the set of control parameters for which the system exhibits multistability [9, 10]: in the case of $\alpha = 1.5$ and $\gamma = 3.0$, the system features oscillations with periods 7, 8, or 15, depending on the initial conditions.

A map of the transient process duration versus the initial conditions was constructed in the Poincaré section of the phase space by the $z = 0$ plane. Since the Poincaré section reduces the $n$-dimensional flow system to $(n - 1)$-dimensional system with discrete time [12], the transient process duration can be determined using the method developed for maps [4, 5]. According to this, the phase trajectory with a length of several iterations is calculated for all values of the initial conditions $(x_0, y_0)$ by the fourth-order Runge–Kutta method with a step of 0.005. This procedure determines a certain sequence of the points of intersection of the phase trajectory with the plane of the Poincaré section. It was suggested a priori that, upon this (sufficiently large) number of iterations, the transient process is completed. Then, the obtained sequence of the intersection points is checked for coincidence of the coordinates of these points to within a preset accuracy. If this verification does not reveal points with coinciding coordinates, a longer transient process is selected and the procedure is repeated until finding a regime attained by the system.[1] The period of a cycle attained by the imaging point is determined by the number of points in the Poincaré section between two points with coinciding coordinates, while the number of iterations accomplished by this moment multiplied by the time step gives the transient process duration. Differences in the transient process duration are mapped in terms of the gray gradation scale.

[1] Algorithm for determining the transient process duration in a chaotic regime is described in [5, 6].

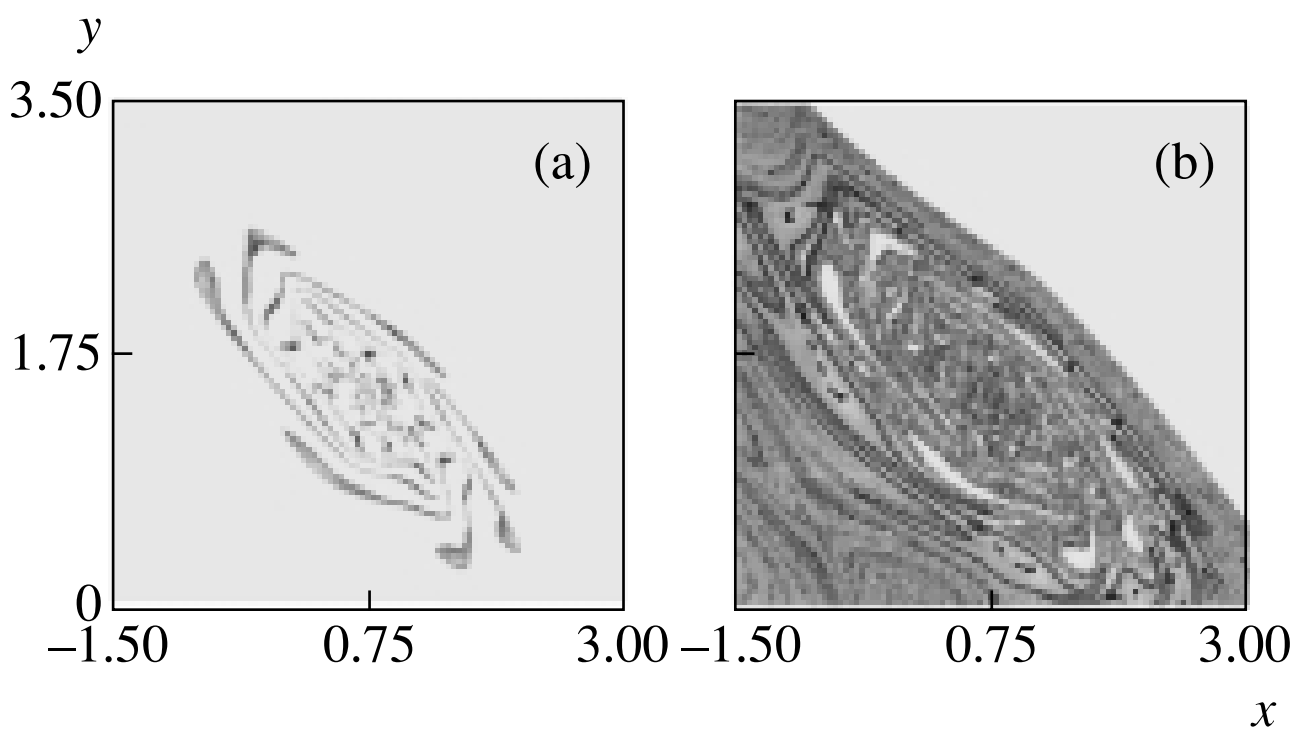

**Fig. 1.** Gray gradation maps of the transient process duration versus initial conditions for the cycles of periods (a) 7 and (b) 8. Bright regions correspond to shorter transient processes, while dark regions represent longer transients. In panel (b): solid and dashed lines indicate stable and unstable manifolds of the saddle cycle 1 : 8, respectively (not all of the unstable manifolds are depicted). Black and bright points correspond to stable and unstable saddle cycles, respectively. Region *1* corresponds to the initial conditions from which the imaging point goes to infinity.

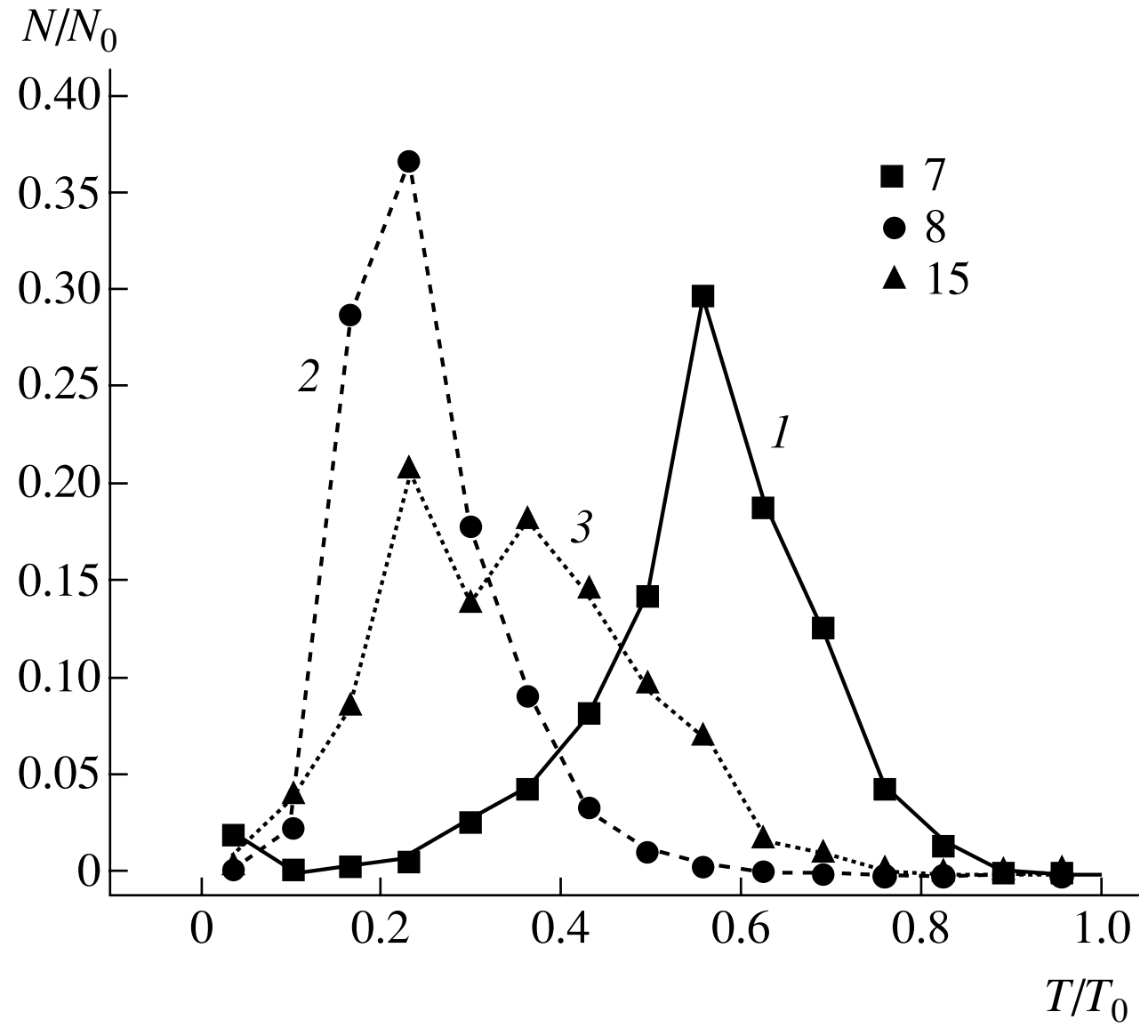

**Fig. 2.** Normalized distributions of the initial points in the basins of attraction with respect to the transient process duration for various cycles. The values in the abscissa axis are normalized to the maximum transient process duration ($T_0$) for the given cycle; the ordinates are normalized to the total number of points ($N_0$) in the given basin of attraction: (*1*) cycle 1 : 7 ($T_0$ = 1519, $N_0$ = 36339); (*2*) cycle 1 : 8 ($T_0$ = 3130, $N_0$ = 420900); (*3*) cycle 2 : 15 ($T_0$ = 3681, $N_0$ = 72198).

Figures 1a and 1b present maps of the transient process duration versus initial conditions for the cycles of period 7 and 8, respectively. As can be seen, these maps exhibit more and less intensely colored regions. Dark regions are situated at the boundaries of shaded areas, while bright regions are situated inside these areas. Shorter transient processes correspond to bright regions, while longer transients are mapped by the dark regions.

We have also determined which transient process durations (longer or shorter) dominate for a given set of control parameters in each oscillation regime. For this purpose, we constructed the distribution of initial conditions ($x_0$, $y_0$) occurring in the Poincaré section (@N) over the intervals of transient process durations ($\Delta T$). To determine this, we select a certain interval [$T$, $T + \Delta T$] and count the number of initial conditions in the basin of attraction in the Poincaré section for which the transient process durations fall within the selected interval. The corresponding normalized distributions are presented in Fig. 2, where values in the abscissa axis are normalized to the maximum transient process duration ($T_0$) and those in the ordinate axis are normalized to the total number of points ($N_0$) in the given basin of attraction. With this normalization, the area under curve is equal to unity.

As can be seen from Fig. 2, the peaks of distributions for the cycles of periods 8 and 15 are shifted toward relatively shorter transient process durations (for each of these cycles, the transient process durations vary within more or less broad range). This shift agrees with the data for 1 : 8 cycle in Fig. 1b, where darkest regions (corresponding to maximum transient process durations) occupy the minimum area. Such behavior is related to the arrangement of stable manifolds of the unstable (saddle) cycles forming the boundaries of the basins of attraction of the corresponding attractors. Proximity of the peak of distribution for the cycle of period 1 : 7 to the middle of the interval of transient process durations can be explained by a relatively small size of the basin of attraction for this cycle in comparison to the basins of attraction for the other cycles (the total number of points, for which this distribution was constructed is 36000; the corresponding number of points for the 1 : 8 cycle is greater by a factor of ten, while the number of points for the 2 : 15 cycle is greater by a factor of two). The main part of the basin of attraction for the 1 : 7 cycle occurs in a region distant from the attracting cycle. This implies that imaging points with the initial conditions distant from the limit cycle spend a rather long time for attaining the attracting cycle.

Another peculiarity observed in Fig. 2 is the presence of two humps in the distribution of initial conditions for the 2 : 15 cycle. This pattern can be also related to certain features in the structure of the given basin of attraction, which is significantly scattered over the entire Poincaré section. Another possible reason for this irregularity could be insufficient statistics, but the results of calculations of the maps of transient process durations with a fourfold difference in the numbers of initial conditions showed no qualitative changes in the shape of distribution. Therefore, we can ascertain that the observed distribution shape is inherent in the system studied.

Figure 1b shows the arrangement of stable and unstable manifolds of the saddle limit cycle 1 : 8 in the Poincaré section. Near the immobile point, the manifold represents a straight segment coinciding with the
1 eigenvector of the monodromy matrix [13]. Arbitrarily selecting a certain number of points in the vicinity of this manifold and tracing their evolution, it is possible to determine the arrangement of manifolds (more rigorous methods for determining this arrangement are described in [14]). Stable and unstable manifolds are indicated by solid and dashed lines, respectively. Intersections of the unstable saddle cycle and the Poincaré section are indicated by bright points, intersections of the stable cycle and the $z = 0$ plane are indicated by black points. As can be seen from this figure, the darkest regions of the map coincide with stable manifolds of the saddle cycle. These very regions correspond to the maximum transient process durations. This is related to the fact that the motion of an imaging point over such manifold takes a long time because the velocity of this motion tends to zero as the point approaches the manifold (in this case, of the saddle cycle). At the same time, the lines of stable manifolds are directed from the regions corresponding to long transient times to the regions of shorter times surrounding the points of intersection of the stable cycle and the Poincaré section. Here, the stable manifold connects the stable and unstable saddle cycles. Whatever close is the imaging point to the unstable cycle or its stable manifold, it will quite rapidly come to the stable cycle. This means that there is only a small number of imaging points starting from the initial conditions situated in the nearest vicinity of the stable manifold and performing actually long transients. This conclusion is confirmed by the fact that shorter transients are predominating in the distribution of the transient process durations (Fig. 2).

To summarize, we have studied transient processes in a third-order radiophysical flow system (an autooscillator of the Chua type), constructed a detailed map of the transient process durations in the plane of the Poincaré section in the region of control parameters corresponding to multistability, and determined distribution of the initial conditions with respect to the transient process duration. Analysis of these maps and distributions revealed important peculiarities in the behavior of transient processes in the basins of attraction of various cycles.

**Acknowledgments.** This study was supported by the Russian Foundation for Basic Research (project nos. 02-02-16351), the program “Universities of Russia: Basic Research” (project UR.01.01.065), the Program of Support for Leading Scientific Schools, and the Federal Program “Integration.”